\documentclass[preprint,12pt]{aastex}

\providecommand{\eg}{e.g.}
\providecommand{\etal}{et~al.}

\received{}
\revised{}
\accepted{}

\slugcomment{To appear in Nov 2004 \aj}

\shorttitle{GRB~021211}
\shortauthors{S.~T. Holland, et~al.}

\begin{document}


\title{GRB 021211 as a Faint Analogue of GRB 990123: Exploring the
Similarities and Differences in the Optical
Afterglows\protect\footnote{Some of the observations reported here
were obtained at the MMT Observatory, a joint facility of the
Smithsonian Institution and the University of Arizona.  Some of the
observations reported here were obtained with the Magellan 2 (Landon
Clay) telescope at the Las Campanas Observatory.  Some of the
observations reported here were taken at the WIYN Observatory, a joint
facility of the University of Wisconsin-Madison, Indiana University,
Yale University, and the National Optical Astronomy Observatories.}}

\author{Stephen~T.~Holland\altaffilmark{2,3,4},
        David~Bersier\altaffilmark{5},
        J.~S. Bloom\altaffilmark{6,7},
        Peter~M.~Garnavich\altaffilmark{4},
        Nelson Caldwell\altaffilmark{6},
        Peter Challis\altaffilmark{6},
        Robert Kirshner\altaffilmark{6},
        Kevin Luhman\altaffilmark{6},
        Brian McLeod\altaffilmark{6}, \&
        K.~Z.~Stanek\altaffilmark{6}}

\altaffiltext{2}{Code 660.1,
                 Goddard Space Flight Centre,
                 Greenbelt, MD 20771--0003,
                 U.S.A.
                 \email{sholland@milkyway.gsfc.nasa.gov}}

\altaffiltext{3}{Universities Space Research Association}

\altaffiltext{4}{Department of Physics,
                 University of Notre Dame,
                 Notre Dame, IN 46556--5670,
                 U.S.A.
                 \email{pgarnavi@miranda.phys.nd.edu}}

\altaffiltext{5}{Space Telescope Science Institute,
                 3700 San Martin Dr.,\
                 Baltimore, MD 21218,
                 U.S.A.
		 \email{bersier@stsci.edu}}

\altaffiltext{6}{Harvard--Smithsonian Centre for Astrophysics,
                 60 Garden St,
                 Cambridge, MA 02138,
                 U.S.A.
                 \email{jbloom@cfa.harvard.edu,
                        kstanek@cfa.harvard.edu,
                        caldwell@cfa.harvard.edu,
                        pchallis@cfa.harvard.edu,
                        kirshner@cfa.harvard.edu,
                        kluhman@cfa.harvard.edu,
                        bcmcleod@cfa.harvard.edu}}

\altaffiltext{7}{Harvard Society of Fellows, 
                 78 Mount Auburn St, 
                 Cambridge, MA 02138,
                 U.S.A.}


\begin{abstract}

     We present $BV\!{R_C}J\!H\!{K_s}$ photometry of the optical
afterglow of the gamma-ray burst \objectname{GRB~021211} taken at the
Magellan, MMT, and WIYN observatories between 0.7 and 50 days after
the burst.  We find an intrinsic spectral slope at optical and
near-infrared wavelengths of $0.69 \pm 0.14$ at 0.87 days.  The
optical decay during the first day is almost identical to that of
\objectname{GRB~990123} except that \objectname{GRB~021211}'s optical
afterglow was intrinsically approximately 38 times fainter and the
transition from the reverse shock to the forward shock may have
occurred earlier than it did for \objectname{GRB~990123}.  We find no
evidence for a jet break or the cooling break passing through optical
frequencies during the first day after the burst.  There is weak
evidence for a break in the $J$-band decay between 0.89 and 1.87 days
which may be due to a jet.  The optical and infrared data are
consistent with a relativistic fireball where the shocked electrons
are in the slow cooling regime and the electron index is $2.3 \pm
0.1$.  The forward shock appears to have been expanding in a
homogeneous ambient medium during the first day after the burst.  Our
analysis suggests that the jet of \objectname{GRB~021211} may have a
small opening angle and that the total gamma-ray energy is likely to
be much less than the canonical value of $1.33 \times 10^{51}$ erg.
If this is the case then it is possible that most of the energy of the
burst is in another form such as a frozen magnetic field, supernova
ejecta, or a second jet component.  The host galaxy of
\objectname{GRB~021211} is subluminous and has a star formation rate
of at least 1 $\mathcal{M}_\sun$ yr$^{-1}$.

\end{abstract}


\keywords{gamma rays: bursts}


\section{Introduction\label{SECTION:intro}}

     The study of gamma-ray bursts (GRBs) took a giant leap forward in
1997 when observations with the Dutch--Italian satellite {\sl
BeppoSAX\/} identified a fading $X$-ray source inside the error box of
\objectname{GRB~970228} \citep{CFH1997}.  This rapid localization of a
counterpart allowed the fading optical afterglow (OA) to be seen
\citep{GGvP1997,vPGG1997}.  Later that year \citet{MDK1997} found a
redshift of 0.835 for \objectname{GRB~970508}, which confirmed the
cosmological distances of GRBs.  Since then the afterglows of several
dozen GRBs have been localized.  In the cases where the afterglows
have been well localized the median redshift is approximately two and
there is no evidence for extra-Galactic reddening of more than
approximately one mag along the line of sight to the afterglow.  Host
galaxies have been found for almost every well-localized GRB
identified \citep{BKD2002}.

     The FREGATE, WXM, and SXC instruments on board the {\sl High
Energy Transient Explorer II\/} ({\sl HETE-II\/}) satellite detected
gamma-ray burst (GRB) \objectname{GRB~021211} (= {\sl HETE\/} trigger
H2493) in the constellation Cancer at 11:18:34.03 UT on 2002~Dec.~11
\citep{CVV2002,CLR2003}.  The burst consisted of a single peak with a
fast rise and exponential decay profile.  The $t_{90}$ durations were
$2.30 \pm 0.52$ seconds at high energies (85--400 keV) and $8.5 \pm
0.5$ seconds at low energies (2--10 keV).  The energy fluence was
$(0.96 \pm 0.03) \times 10^{-6}$~erg~cm$^{-2}$ between 7 and 30~keV,
and $(1.98 \pm 0.15) \times 10^{-6}$~erg~cm$^{-2}$ between 30 and
400~keV giving a hardness ratio of $S_X / S_\gamma = 0.48$
\citep{CLR2003}.  The gamma-ray properties of \objectname{GRB~021211}
were not unusual and fit into the long--soft class of bursts
\citep{KMF1993}.  The burst was $X$-ray rich with a spectrum that
peaked at $E_\mathrm{peak} = 58$ keV \citep{CLR2003}.

     The field of \objectname{GRB~021211} was observed starting 20.74
minutes after the {\sl HETE-II\/} trigger with the Near Earth Asteroid
Tracking Camera on the Oschin 48-inch telescope at the Palomar
Observatory.  These observations led to the announcement of the
discovery of the OA \citep{FPS2003}.  Observations taken even earlier
by \citet{WVS2002,LFC2003}, and \citet{PWB2002} show that the OA faded
rapidly, with a power-law slope of approximately 1.6, between
approximately 90 seconds and 11 minutes after the burst.  After this
it faded more slowly, with a power-law slope of 0.82 \citep{LFC2003}.
\citet{DBM2003} found a redshift of $z = 1.004 \pm 0.002$ for the host
galaxy based on four emission lines from the host galaxy.

     \citet{W2003} finds that the optical decay of \objectname{GRB
021211} can be explained if the emission before eleven minutes is due
to the reverse shock while the emission after that is due to the
forward shock.  They find that the magnetic field strength is
unusually low and propose that this is why \objectname{GRB~021211}'s
OA was faint.  However, \citet{KP2003} and \citet{PK2004} analyse
radio and optical data for \objectname{GRB~021211} and find that the
observations before eleven minutes after the burst are consistent with
a reverse shock if the energy density in the magnetic field in the
reverse shock is $\epsilon_{B_r} \approx 0.1$, higher than usual for a
GRB\@.  They also find that the local environment around the
progenitor is best fit by a homogeneous interstellar medium (ISM) with
a low particle density ($n_0 \approx 10^{-2}$ to $10^{-3}$ cm$^{-3}$).
Their model suggests that the magnetic field was frozen in during the
explosion, and that the magnetic field contains a significant amount
of the energy from the explosion.

     \citet{DMB2003} used deep imaging and spectroscopy to find
evidence of a supernova component 27 days after the burst.  They find
that the optical spectrum is consistent with that of
\objectname{SN1998bw} and the Type Ic supernova \objectname{SN1994I}.
This made \objectname{GRB~021211} the second GRB---after
\objectname{GRB~030329}/\objectname{SN2003dh}
\citep{SMG2003,MGS2003,HSM2003}---for which spectroscopic confirmation
of a link between GRBs and supernovae was obtained.  \citet{DMB2003}
find $R = 25.22 \pm 0.10$ mag for the host galaxy.

     In this paper we adopt a cosmology with a Hubble parameter of
$H_0 = 70$~km~s$^{-1}$~Mpc$^{-1}$, a matter density of $\Omega_m =
0.3$, and a cosmological constant of $\Omega_\Lambda = 0.7$.  For this
cosmology a redshift of $z = 1.004$ corresponds to a luminosity
distance of 6.64~Gpc and a distance modulus of 44.11.  One arcsecond
corresponds to 16.06 comoving kpc, or 8.02 proper kpc.  The look back
time is 7.75~Gyr.


\section{The Data\label{SECTION:data}}

     The OA for \objectname{GRB~021211} is located at R.A. $=
08^\mathrm{h} 08^\mathrm{m} 59\fs858$, Dec.\ $= +06\arcdeg 43\arcmin
37\farcs52$ (J2000) \citep{FPS2003}, which corresponds to Galactic
coordinates of $(b^\mathrm{II},l^\mathrm{II}) =
(+20\fdg2950,\linebreak[0] 215\fdg7486)$.  The reddening maps of
\citet{SFD1998} give a Galactic reddening of $E_{B-V} = 0.028 \pm
0.020$~mag in this direction.  The corresponding Galactic extinctions
are $A_U = 0.149$, $A_B = 0.119$, $A_V = 0.091$, $A_{R_C} = 0.074$,
$A_{I_C} = 0.053$, $A_J = 0.026$, $A_H = 0.016$, and $A_K = 0.010$
mag.


\subsection{Optical Photometry\label{SECTION:optical}}

     Optical imaging data were obtained on 2002 Dec.\ 12 UT using the
Minicam on the 6.5m Multi-Mirror Telescope (MMT) at the MMT
Observatory.  The instrumental gain was 2.7 e$^-$ ADU$^{-1}$, the
readout noise was 4.5 e$^-$ per pixel, and the plate scale was
$0\farcs091$ per pixel.

     We obtained additional optical imaging data on 2003 Jan.\ 2 UT
using the Low Dispersion Survey Spectrograph 2 (LDSS-2) in its imaging
mode on the 6.5m Magellan 2 Landon Clay telescope at the Las Campanas
Observatory.  The gain was 3.8 e$^-$ ADU$^{-1}$, the readout noise was
10.2 e$^-$ per pixel, and the plate scale was $0\farcs378$ per pixel.

     Our final set of optical imaging data were obtained on 2003 Jan.\
29 UT using Minimosaic on the WIYN 3.5m telescope at Kitt Peak.  The
gain was 1.5 e$^-$ ADU$^{-1}$, the readout noise was 5.5 e$^-$ per
pixel, and the plate scale was $0\farcs141$ per pixel.

\begin{figure}
\plotone{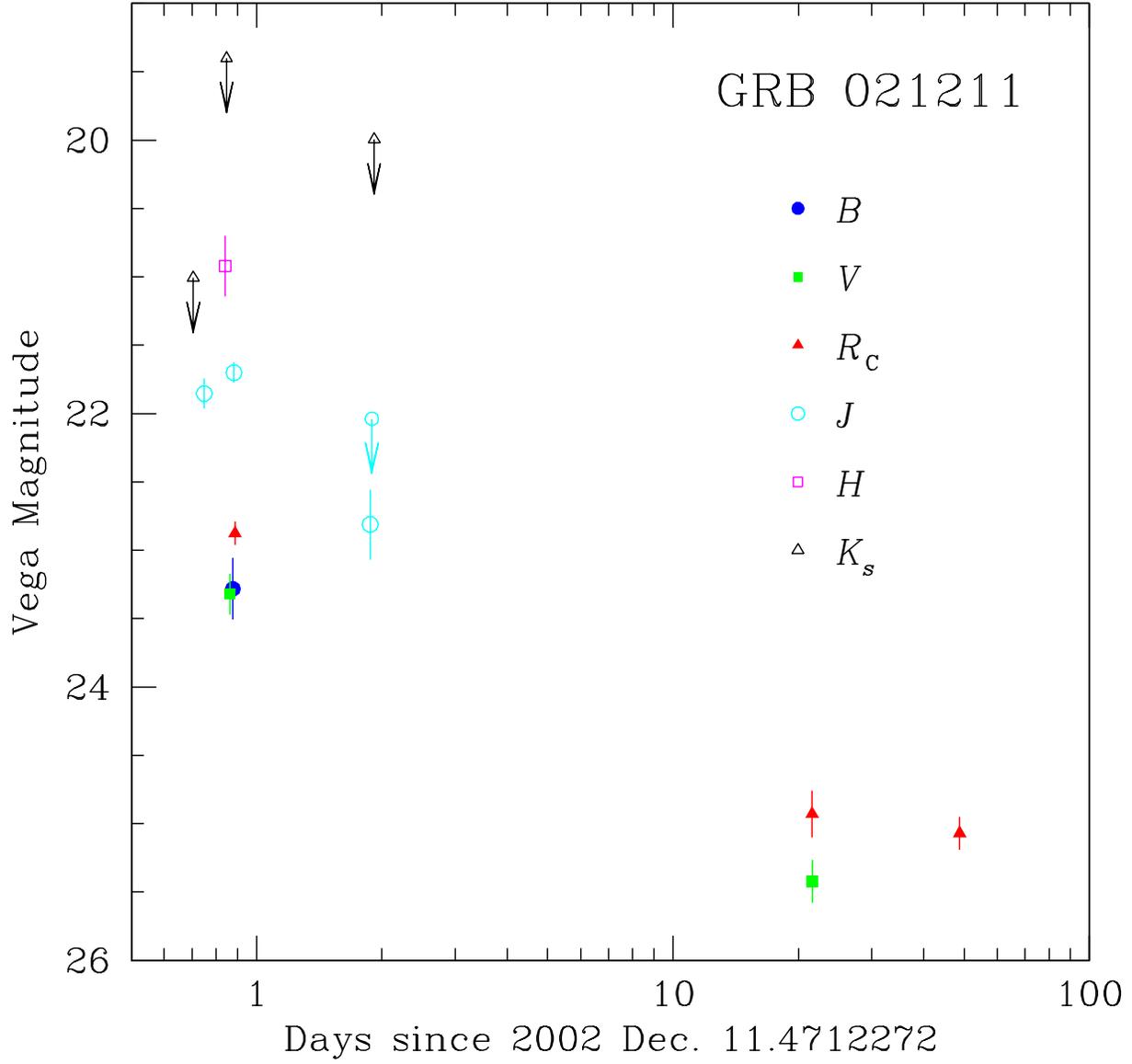} \figcaption[./our_data.eps]{These are our
optical and infrared data for \protect\objectname{GRB~021211}.  The
closed circles are $B$-, the closed squares are $V$-, the closed
triangles are $R_C$-, the open circles are $J$-, the open squares are
$H$-, and the open triangles are $K_s$-band data.  The arrows indicate
upper limits.\label{FIGURE:our_data}}
\end{figure}

     A log of our observations and the photometry of the OA is
presented in Table~\ref{TABLE:phot}.  No reddening corrections have
been applied to the data in this Table.  Figure~\ref{FIGURE:our_data}
shows the optical data presented in this paper.

\begin{deluxetable}{lcrrcccr}
\tabletypesize{\scriptsize}
\tablewidth{0pt}
\tablecaption{Log of the \protect\objectname{GRB~021211} observations
and the results of the photometry.  The UT date is the middle of each
exposure.\label{TABLE:phot}}
\tablehead{%
     \colhead{UT Date} &
     \colhead{JD $-$ 2450000} &
     \colhead{$t$ (days)} &
     \colhead{Telescope} &
     \colhead{Filter} &
     \colhead{Magnitude} &
     \colhead{Seeing ($\arcsec$)} &
     \colhead{Exptime (s)}}
\startdata                
 Dec.\ 12.17617 & 2620.67617 &  0.70494 & Magellan 1/ClassicCam & $K_s$ & $>21.01$           & $0\farcs44$ &  660 \\
 Dec.\ 12.22002 & 2620.72002 &  0.74879 & Magellan 1/ClassicCam & $J$   &   21.85 $\pm$ 0.11 & $0\farcs48$ & 3720 \\
 Dec.\ 12.31163 & 2620.81163 &  0.84040 & Magellan 1/ClassicCam & $H$   &   20.92 $\pm$ 0.22 & $0\farcs35$ &  240 \\
 Dec.\ 12.31873 & 2620.81873 &  0.84750 & Magellan 1/ClassicCam & $K_s$ & $>19.4$            & $0\farcs49$ &  240 \\
 Dec.\ 12.33329 & 2620.83329 &  0.86206 & MMT/Minicam           & $V$   &   23.32 $\pm$ 0.15 & $0\farcs76$ & 500 \\
 Dec.\ 12.34816 & 2620.84816 &  0.87693 & MMT/Minicam           & $B$   &   23.28 $\pm$ 0.23 & $0\farcs66$ & 900 \\
 Dec.\ 12.35336 & 2620.85336 &  0.88213 & Magellan 1/ClassicCam & $J$   &   21.70 $\pm$ 0.08 & $0\farcs40$ & 4680 \\
 Dec.\ 12.35917 & 2620.85917 &  0.88794 & MMT/Minicam           & $R_C$ &   22.88 $\pm$ 0.09 & $0\farcs51$ & 900 \\
 Dec.\ 13.34512 & 2621.84512 &  1.87389 & Magellan 1/ClassicCam & $J$   &   22.81 $\pm$ 0.26 & $0\farcs45$ & 4320 \\
 Dec.\ 13.36368 & 2621.86368 &  1.89245 & Magellan 1/ClassicCam & $J$   & $>22.04$           & $0\farcs41$ & 2400 \\
 Dec.\ 13.38875 & 2621.88875 &  1.91752 & Magellan 1/ClassicCam & $K_s$ & $>20.00$           & $0\farcs42$ &  210 \\
 Jan.\  2.06488 & 2641.56488 & 21.59365 & Magellan 2/LDSS-2     & $R_C$ &   24.93 $\pm$ 0.17 & $1\farcs13$ & 300 \\ 
 Jan.\  2.08051 & 2641.58051 & 21.60928 & Magellan 2/LDSS-2     & $V$   &   25.42 $\pm$ 0.16 & $0\farcs75$ & 300 \\ 
 Jan.  29.27962 & 2668.77962 & 48.80839 & WIYN/Minimosaic       & $R_C$ &   25.07 $\pm$ 0.12 & $0\farcs58$ & 600 \\
\enddata
\end{deluxetable}

     The data were preprocessed using standard techniques for bias and
flat-field corrections.  Photometry was performed by fitting a
point-spread function using the {\sc DAOPhot~II} and {\sc AllStar}
software \citep{S1987,SH1988}.  The MMT and Magellan data were
calibrated using secondary standard stars from the catalogue of
\citet{H2002}.  Each image was calibrated using all of the
\citet{H2002} secondary standards that could be photometered on that
image.  A weighted mean of the calibrated magnitudes of each of our
detections of the secondary standards was used to determine the
photometric zero point for each image.  The colour terms in the
calibrations were smaller than 5\% and did not improve the quality of
the calibrations.  Therefore they were not applied to the photometry.
The WIYN image from 2003 Jan.\ 29 was very deep, so the secondary
standards of \citet{H2002} that were in the field of view were
saturated.  Therefore the zero point for this image was determined by
matching stars on the $R_C$-band image obtained with the MMT on 2003
Dec.\ 12.


\subsection{Infrared Photometry\label{SECTION:ir}}

     Infrared imaging data were obtained on 2003 Dec.\ 12, and 13 UT
using the Classic Camera's NICMOS3 detector on the 6.5m Magellan 1
(Walter Baade) telescope at the Las Campanas Observatory.  The gain
was 7.5 e$^-$ ADU$^{-1}$, the readout noise was 40 e$^-$ per pixel,
and the plate scale was $0\farcs095$ per pixel.  A dithering sequence
with steps of a few arcseconds was used for the individual images.
The data were preprocessed in the standard way and the individual
images were coadded to produce a final mean image for each epoch.  The
brightest and faintest $\approx 5$\% of the pixels at each location in
the stack were rejected in order to eliminate cosmic rays and bad
pixels that were missed in the preprocessing steps.

     Several faint infrared standard stars from the lists of
\citet{PMK1998} were observed on each night to determine the
photometric zero points of our images.
The standard star data were preprocessed in exactly the same manner as
the programme data, and aperture photometry was performed on the
individual images.  The zero-point offset for each standard was
computed for each observation of that standard, and a weighted mean
zero point was computed from all of the standards in each filter on
each night.  We found no evidence for significant time evolution in
the zero point over the course of a night.  Colour terms do not
improve the quality of the calibration, so they were not used.

     A log of our infrared observations and the photometry of the OA
is included in Table~\ref{TABLE:phot}.  No reddening corrections have
been applied to this data.  Figure~\ref{FIGURE:our_data} shows the
infrared data presented in this paper.


\section{Results\label{SECTION:results}}

\subsection{The Optical Light Curve\label{SECTION:light_curve}}

\begin{figure}
\plotone{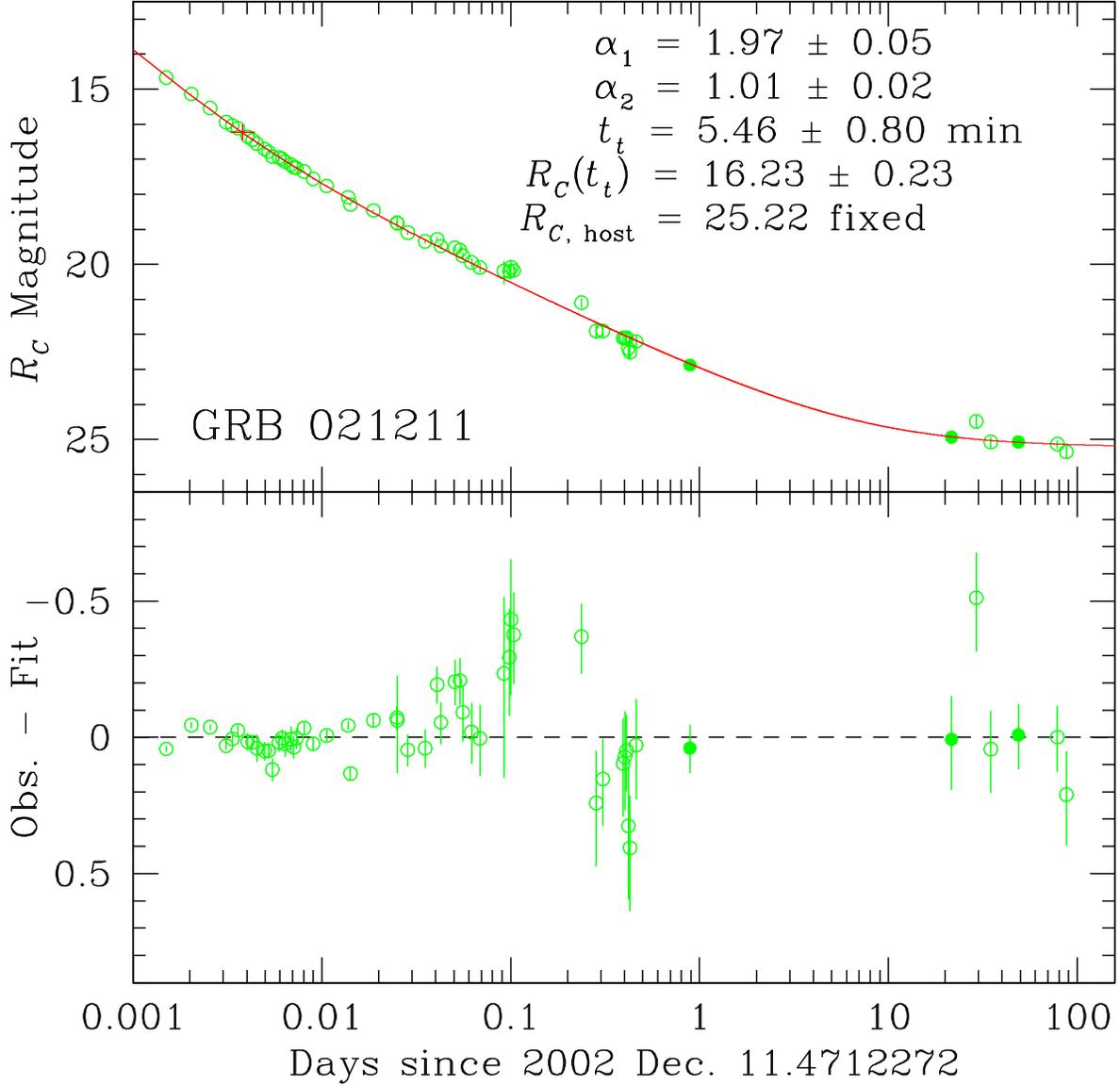} \figcaption[./decay.eps]{The upper panel shows
our $R_C$-band data and the published data for
\protect\objectname{GRB~021211}.  The open circles represent the
published data while the solid circles represent our data.  The line
is the best-fitting double power law to the $R_C$-band data for $t \le
1$ days as described in \S~\ref{SECTION:light_curve}.  The cross at
the location of the break shows the size of the $1\sigma$ error bars
in the fitted break time.  The lower panel shows the residuals in the
fit.  The variations during the first day are
real.\label{FIGURE:decay}}
\end{figure}

     To parameterize the optical decay we combined our $R_C$-band data
with the published data from \citet{FPS2003}, \citet{LFC2003},
\citet{PAS2003}, and \citet{DMB2003} (see Figure~\ref{FIGURE:decay}).
We fit the data with a double power law of the form

\begin{equation}
  \label{EQUATION:double_power_law}
     R_{C,\mathrm{fit}}(t) =
        -48.77
	   - 2.5 \log_{10}\left( 0.5 f_{\nu}(t_t)
	      \left[ {(t/t_t)}^{-\alpha_1}
              +  {(t/t_t)}^{-\alpha_2} \right] \right)
           + R_{C,\mathrm{host}},
\end{equation}

\noindent
where $t_t$ is the time of the transition from the first to the second
power law component, $f_{\nu}(t_t)$ is the flux density in the $R_C$
band at the time of the transition, $R_{C,\mathrm{host}}$ is the
magnitude of the host galaxy, and $\alpha_1,\alpha_2$ are the decay
slopes before and after the transition respectively.  The constant
$-48.77$ converts flux density to magnitude and is based on the zero
points of \citet{FSI1995}.  This formalism assumes that there are two
components to the light from the OA\@.  Both components contribute to
the total light and each component fades at different rates.  We fit a
double power law instead of a broken power law because the preferred
interpretation of the early behaviour of the OA of
\objectname{GRB~021211} is a reverse shock.  Therefore, the OA will
contain contributions from both the reverse and forward shocks with
the transition time indicating when the contribution from each was
equal.  A broken power law assumes that there was a sharp transition
between the two decay rates.  This is (approximately) valid for the
transition from a relativistically beamed jet to a
non-relativistically beamed jet, but it is not valid for the
transition between the OA being dominated by a reverse and forward
shock.

     We fit Eq.~\ref{EQUATION:double_power_law} to all of the
$R_C$-band data for the first day.  We did not include the supernova
component in our fit since its contribution is negligible at these
early times.  The magnitude of the host galaxy was fixed at
$R_{C,\mathrm{host}} = 25.22$ mag \citep{DMB2003}.  Our best fit is
shown in Figure~\ref{FIGURE:decay}.  We find power law slopes of
$\alpha_1 = 1.97 \pm 0.05$ and $\alpha_2 = 1.01 \pm 0.02$; a
transition time of $t_t = 5.46 \pm 0.80$ minutes; and a flux density
at the transition of $f_{\nu}(t_t) = 0.996 \pm 0.214$ mJy, which
corresponds to $R_C(t_t) = 16.12 \pm 0.23$ mag.  The fit does not
change if we allow the host magnitude to be a free parameter.  In this
case the best fit for $R_{C,\mathrm{host}}$ is $25.25 \pm 0.07$ mag.
Our fit is formally not good.  The chi-square value is 162.1 for 45
degrees of freedom.  Reducing this to $\chi^2/45 = 1$ would require
that we have underestimated the errors in the magnitudes by a factor
of approximately two.  Since we have used multiple data sets it is
possible that cross-calibration errors may be present.  However, in
Sect~\ref{SECTION:fluctuations} we show that the large residuals are
present in fits to individual data sets, so the large chi-square value
that we find is not due cross-calibration errors.

     The double power law does not include the effects of the
synchrotron peak frequency, $\nu_m$, moving through the optical.  The
is expected to happen at early times ($t \approx 5$--15 minutes).  If
$\nu_m$ did pass through the $R_C$ band at approximately the
transition time between the forward and reverse shocks it will result
in the best-fit slopes and transition time in
Eq.~\ref{EQUATION:double_power_law} being incorrect.

     Our late-time Magellan data is consistent with there being no
variation in the $R_C$-band magnitude of the host galaxy between 22
and 49 days after the burst.  The only evidence for a supernova bump
in the optical decay is the photometry of \citet{DMB2003}, who find
$R_C = 24.48 \pm 0.18$ mag at 29 days after the burst, which is
approximately 0.5 mag ($\approx 2.6 \sigma$) brighter than the
combined OA and host galaxy at that time.  It should be noted,
however, that the OA of \objectname{GRB 030329} did not exhibit a
strong rebrightening when \objectname{SN2003dh} was near its peak
brightness.  It is possible that the late-time OA of
\objectname{GRB~021211} behaved in a similar manner.  This suggests
that the lack of an optical bump may not indicate the lack of a
supernova.

     We find no evidence for a second break in the optical decay
during the first day.  Our $J$-band photometry suggests that the decay
between 21.2 and 45 hours was $\alpha = 1.36 \pm 0.33$.  This is
steeper than the decay in the $R_C$ band during the first day and may
indicate that a break occurred during this time.  In
\S~\ref{SECTION:environment} we argue that the cooling break is most
likely to be at $X$-ray frequencies during this time, so we consider
it very unlikely that this apparent steepening of the decay is due to
the cooling frequency passing through the $J$ band.  Therefore it is
possible that we are seeing the jet break.  However, the $J$ band flux
density increased by $\approx 15$\% between 18 and 21.2 hours, so it
is also possible that we are seeing a short-term fluctuation in the
$J$-band flux density (see \S~\ref{SECTION:fluctuations}) and not a
break in the decay.


\subsection{The Spectral Energy Distribution\label{SECTION:sed}}
         
     We combined our photometry with data drawn from the literature
\citep{FPS2003,LFC2003,PAS2003} to construct spectral energy
distributions (SEDs) for the OA at several epochs.  The optical and
infrared magnitudes were converted to flux densities based on the zero
points of \citet{FSI1995} and \citet{A2000}.  Each data point was
corrected for Galactic reddening but not for any reddening that may be
present in the host galaxy, or in intergalactic space between us and
the host.
     
     The SED was fit by $f_{\nu}(\nu) \propto \nu^{-\beta} \times
10^{-0.4 A(\nu)}$, where $f_\nu(\nu)$ is the flux density at frequency
$\nu$, $\beta$ is the intrinsic spectral index, and $A(\nu)$ is the
extragalactic extinction along the line of sight to the burst.  The
dependence of $A(\nu)$ on $\nu$ has been parameterized in terms of the
extinction in the rest-frame $B$ band, $A_B$.  We adopted the three
extinction laws given by \citet{P1992}: the Milky Way (MW), the Large
Magellanic Cloud (LMC), and the Small Magellanic Cloud (SMC).  The fit
provides $\beta$ and $A_B$ simultaneously for each of the assumed
extinction laws.  The unextincted case ($A_B = A_V \equiv 0$ mag) was
also considered.

     There are enough data to estimate the amount of extinction only
at 0.87 days after the burst.  We find that our data are consistent
with $\beta = 0.69^{+0.14}_{-0.14}$ and $A_V = 0.0^{+0.08}_{-0.00}$
mag in the host galaxy ($\chi^2/\mathrm{DOF} = 0.5446$).  The three
extinction laws of \citet{P1992} do not provide a statistically better
fit for any value of $A_V$ in the host than $A_V = 0$ mag does.  In
addition, the lack of curvature in the spectrum after correction for
Galactic extinction implies that there is very little extinction in
the host galaxy along the line of sight to \objectname{GRB~021211}.
Therefore we believe that there is negligible extinction affecting
this OA in its host galaxy.

     To test for dust along the line of sight between us and the host
we repeated our fits allowing the redshift of the dust to be a free
parameter.  In these cases the best fits for each extinction law were
for no dust at any redshift in the interval $0 \le z \le 1.004$.
Therefore, we conclude that the dim nature of the OA of
\objectname{GRB 021211} must be intrinsic to the burst itself and not
the result of absorption along the line of sight to the burst.

\begin{figure}

\plotone{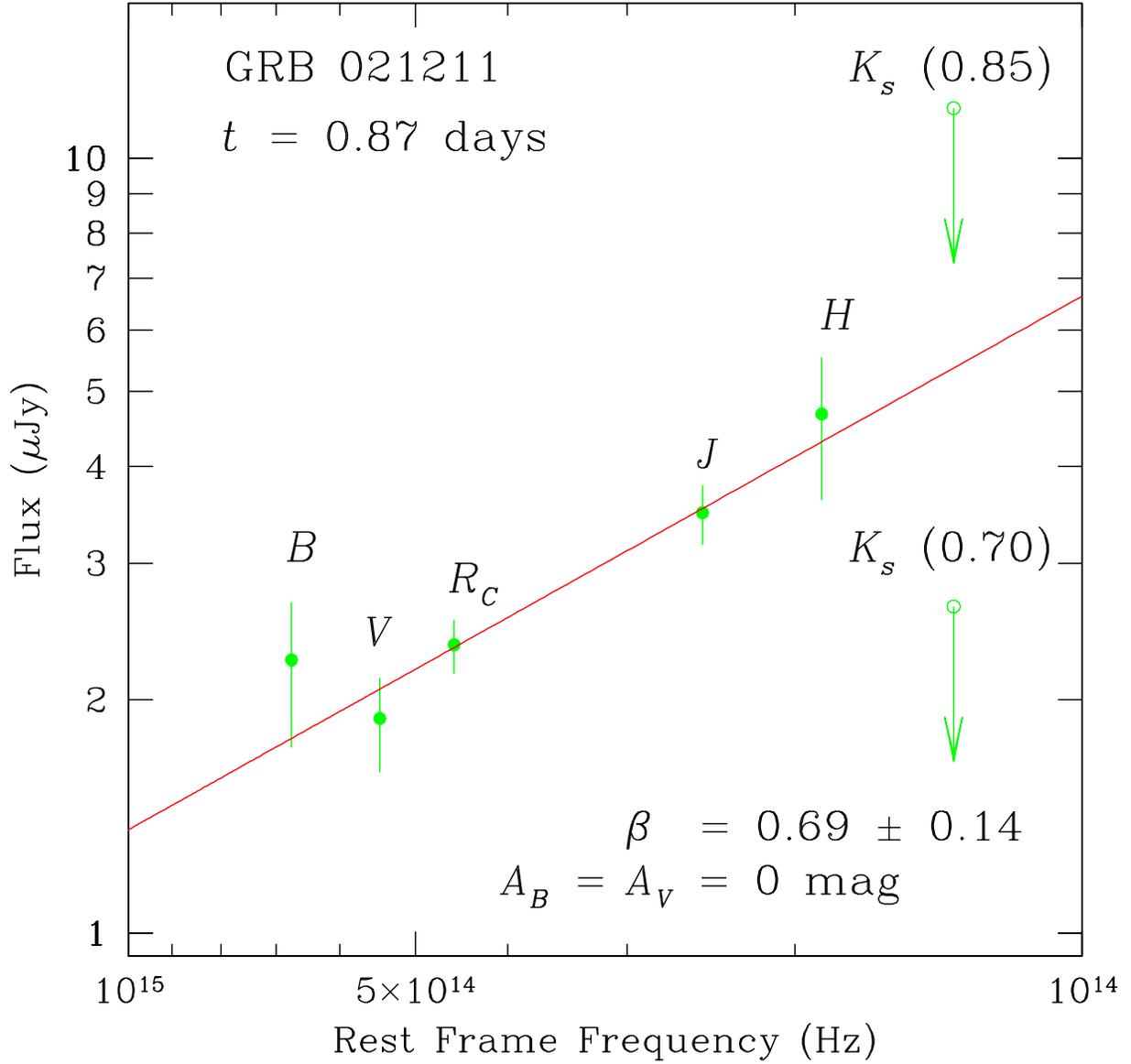} \figcaption[./sed.eps]{The SED of the OA of
\protect\objectname{GRB~021211} on 2002 Dec.\ 12.341 UT\@.  The filled
circles represent observed photometry corrected for extinction in the
Milky Way.  The line represents the SED fit assuming no extinction of
the host.  If we assume that the unextincted spectrum follows
$f_\nu(\nu) \propto \nu^{-\beta}$ then the best fit has $\beta =
0.69^{+0.14}_{-0.14} \pm 0.14$ and $A_B = A_V = 0^{+0.08}_{-0.00}$
mag.  The $K_s$ band upper limits at approximately 0.87 days are shown
as open circles with the time, in days since the burst, given in
parentheses.  Note that the predicted $K_s$-band flux density at 0.87
days (5.367 $\mu$Jy, $K_s = 20.23$ mag) is greater than the upper
limit on the $K_s$-band flux density at 0.70 days (2.638 $\mu$Jy, $K_s
= 21.00$ mag) which implies that the infrared afterglow became
brighter between 0.7 and 0.87 days.
\label{FIGURE:sed}}

\end{figure}


\subsection{Fluctuations in the Decay\label{SECTION:fluctuations}}

      Rapid variability has been seen in the early optical decay of
several GRB afterglows. \objectname{GRB~011211} showed fluctuation of
approximately 15\% over a period of 2.5 hours approximately twelve
hours after the burst \citep{JHR2004,HSG2002}.
\objectname{GRB~021004} exhibited rapid variations in flux density and
colour approximately one day after the burst \citep{BSW2003,MGF2003}.
\citet{MGS2003} have also found rapid colour variations in the OA of
\objectname{GRB~030329} approximately 0.75 days after the burst.
Figure~\ref{FIGURE:decay} suggests that there may be small-scale
variability in the $R_C$-band decay of \objectname{GRB~021211}.  The
anomalous data point at 29 days may be due to light from a supernova
\citep{DBM2003}, so we will only consider the optical decay before
approximately one day after the burst.  The residuals about our
best-fitting double power law between approximately one minute and one
day after the burst are not consistent with zero.  We find that the
$\chi^2$ probability of observing these residuals about a smooth decay
by chance is $2 \times 10^{-14}$.  This is sufficient to rule out a
smooth decay between approximately 1.5 minutes and one day after the
burst.  A synchrotron break passing through the $R_C$ band can produce
a bump in the light curve similar to those seen during this time.
However, it can not produce the multiple bumps that are observed.  We
can not rule out the possibility that some of the variation seen in
the optical decay is due to the passage of the synchrotron break.

     To test that this result is not due to a single data set, which
would suggest that the fluctuations are an artifact of the data
reduction, we computed the $\chi^2$ probability of observing the
residuals about a smooth decay by chance for each data set.  The
\citet{FPS2003} data have a probability that the residuals were due to
chance of $3 \times 10^{-9}$.  The \citet{LFC2003} data have a
probability that the residuals were due to chance of $1 \times
10^{-8}$.  The \citet{PAS2003} data are consistent with random
fluctuations about a smooth decay.  It is not possible to determine if
the systematic offset of +0.14 mag from the best-fitting smooth decay
seen in the \citet{PAS2003} data is real or an artifact of their data
reduction.  The offset is consistent with the fluctuations seen in the
other data sets, so we have no reason to believe that it is not real.

     The upper limit on the $K_s$-band flux density at 0.70 days is
2.638 $\mu$Jy ($K_s = 21.00$ mag).  This is 2.729 $\mu$Jy (0.77 mag)
fainter than the flux density predicted by extending the fitted
spectrum at 0.87 days to the $K_s$ band.  This suggests that the
infrared afterglow has brightened during this time.  The $J$-band flux
density also appears to have brightened during this time.  If the
increase in the $J$-band flux density is part of an achromatic
fluctuation in the flux density then it predicts $K_s = 20.43$ mag at
0.705 days.  However, the upper limit is $K_s \ge 21.00$ mag at this
time.  This suggests that the infrared fluctuations are {\sl not\/}
achromatic.

     \citet{PAS2003} report spectral slopes, corrected for Galactic
reddening, at three epochs between 0.13 and 0.86 days after the burst.
Their slopes are consistent with a constant spectral slope during this
time of $\beta = 0.56 \pm 0.13$ ($B\!-\!V = 0.13$ mag) at the 93\%
confidence level.  At $t = 0.1$ days \citet{FPS2003} find $\beta =
0.98$ ($B\!-\!V = 0.23$) after correcting for Galactic reddening.  We
find an intrinsic spectral slope of $\beta = 0.69 \pm 0.14$ mag ($B\!-\!V
= 0.28$ mag) at $t = 0.87$ days.  The only evidence for a change in colour
are the data of \citet{FPS2003} who derived their spectral slope from
$B$- and $K_s$-band photometry.  When we combine our colour data with
that in the literature \citep{FPS2003,LFC2003,PAS2003} we find no
evidence for a change in colour between 0.1 and 1.0 days.  However, we
note that there is too little published data to be able to make any
strong statements about colour variations during this time.
   

\section{Physics of the Burst\label{physics}}

\subsection{Energy Considerations\label{energy}}

     We computed the isotropic equivalent energy of
\objectname{GRB~021211} using the observed {\sl HETE-II\/} fluences
listed in \citet{CLR2003}.  We applied a cosmological $K$ correction
\citep{BFS2001} to the energy in each band in order to correct it to
the 20--2000 keV band then averaged the six results.  The resulting
$K$-corrected isotropic equivalent energy is $E_\mathrm{iso} = (1.0
\pm 0.1) \times 10^{52}$ erg.

     We find no evidence for a break in the optical decay due to
beaming before approximately one day after the burst.  However,
\citet{KP2003} note that the radio flux density at 8.5 GHz at 10 days
is a factor of 5.4 lower than expected.  This discrepancy can be
explained if a jet break occurred between one and ten days after the
burst.  If the steepening of the $J$-band decay that we observe
between 21.2 and 45 hours after the burst is due to a jet break then
the time of the break is constrained to 0.89 and 1.87 days.

     The opening angle of a GRB jet can be estimated using
\citet{R1999} and \citet{SPH1999}.  This requires a knowledge of the
circumburst density, which is not well understood for
\objectname{GRB~021211}.  Detailed modeling by \citet{KP2003} implies
$10^{-3} \la n \la 10^{-2}$ cm$^{-3}$.  If we assume an efficiency of
$\eta_\gamma = 0.2$ and a jet break between one and ten days the
opening half-angle of the jet is $1\fdg4 \la \theta_j \la 4\fdg4$.  If
the \citet{KP2003} density range is correct then
\objectname{GRB~021211}'s jet has one of the smallest known
half-opening angles.  Even if the circumburst density is $\approx 100$
cm$^{-3}$ the corresponding jet has $\theta_j \approx 14\arcdeg$ at
ten day after the burst.  The beaming-corrected energies which
correspond to the estimated range of jet angles are $3 \times 10^{48}$
erg for $\theta_j = 1\fdg4$ and $t_j = 1$ day, and $2.9 \times
10^{50}$ erg for $\theta_j = 14\arcdeg$ and $t_j = 10$ days.  These
energies are approximately 5 to 400 smaller than the canonical value
of \citet{BFK2003}.  In order for \objectname{GRB~021211} to have the
canonical energy the jet half-opening angle needs to be $\approx
30\arcdeg$.  If this is correct then either the jet break will not
occur until $t \gg 10$ days after the burst or the ambient density
exceeds $\approx 48\,000$ cm$^{-3}$.  Both of these scenarios are
unlikely so we believe that \objectname{GRB~021211} may have been
underluminous at gamma-ray energies.


\subsection{The Cooling Frequency and Local Environment of
GRB~021211\label{SECTION:environment}}

     Most GRBs are well fit by relativistic fireball models with an
electron index of $p \approx 2.3$--2.5 \citep{VKW2000}.
\citet{PK2002} present models for ten GRBs and find a mean electron
index of $\overline{p} = 1.9$ with five GRBs having $p < 2$.  This is
problematic since electron indices of less than two represent infinite
energy in the standard relativistic fireball model \citep{M2002}.
This problem can be avoided by introducing an upper limit for the
electron energy distribution \citep{DC2001}.  Detailed modeling of the
acceleration of particles in highly relativistic shocks predict an
electron index of approximately 2.3 \citep{AGK2001}.

     We estimate the electron index for \objectname{GRB~021211} using
the slope of the optical decay, and the spectral index, of
\objectname{GRB~021211} at 0.87 days.  The optical decay has a slope
of $\alpha = 1.01 \pm 0.02$ and the dereddened slope of the SED at
optical wavelengths is $\beta = 0.69 \pm 0.14$ ($B\!-\!V = 0.28 \pm
0.03$).  We used the relationships of \citet{SPN1998} (for a
homogeneous interstellar medium (ISM)) and \citet{CL1999} (for a
pre-existing stellar wind) to predict $p$ at this time.  These
predictions are listed in Table~\ref{TABLE:predict} for several stages
of the evolution of a synchrotron emission spectrum.

\begin{deluxetable}{cclcll}
\tabletypesize{\scriptsize}
\tablewidth{0pt}
\tablecaption{Predicted electron and spectral indices between 12.1
minutes and 21.3 hours after the burst assuming $\alpha = 1.01 \pm
0.02$.  The electron index is predicted from the observed decay.  The
spectral index is derived from the predicted value of
$p$.\label{TABLE:predict}}
\tablehead{%
        \colhead{Case} &
        \colhead{Model} &
        \colhead{Environment} &
        \colhead{$p$} &
        \colhead{$\beta$} &
        \colhead{Comments}}
\startdata
  1 & $\nu_m < \nu_c < \nu$ & ISM  & $2.0 \pm 0.1$ & $1.01 \pm 0.01$ & \\
  2 &                       & Wind & $2.0 \pm 0.1$ & $1.01 \pm 0.01$ & \\
  3 & $\nu_m < \nu < \nu_c$ & ISM  & $2.3 \pm 0.1$ & $0.67 \pm 0.01$ & \\
  4 &                       & Wind & $1.7 \pm 0.1$ & $0.34 \pm 0.01$ & \\
  5 & $\nu < \nu_m < \nu_c$ & ISM  &    \nodata    & $-0.33$         & Rising spectrum \\
  6 &                       & Wind &    \nodata    & $-0.33$         & Rising spectrum \\
  7 & $\nu_c < \nu_m < \nu$ & ISM  & $2.0 \pm 0.1$ & $1.01 \pm 0.01$ & \\
  8 &                       & Wind & $2.0 \pm 0.1$ & $1.01 \pm 0.01$ & \\
  9 & $\nu_c < \nu < \nu_m$ & ISM  &    \nodata    & $0.5$           & $\alpha = 0.25$ \\
 10 &                       & Wind &    \nodata    & $0.5$           & $\alpha = 0.25$ \\
 11 & $\nu < \nu_c < \nu_m$ & ISM  &    \nodata    & $-0.33$         & Rising spectrum \\
 12 &                       & Wind &    \nodata    & $-0.33$         & Rising spectrum \\
\enddata
\end{deluxetable}

     Cases 5, 6, 11, and 12 can be ruled out because they predict a
spectrum with an opposite slope to what is observed.  Cases 9 and 10
can be ruled out because they predict that the optical decay has a
slope of $\alpha = 0.25$, which is not consistent with the observed
slope.  The SED at $t = 0.87$ days is barely consistent ($\approx 2.5
\sigma$) with a prediction of $\beta = 1.01$ (cases 1, 2, 7, and 8)
and $\beta = 0.34$ (case 4).  It is, however in good agreement with
$\beta = 0.67$ (case 3).  This suggests that \objectname{GRB~021211}
was in the slow cooling regime with the synchrotron frequency,
$\nu_m$, below the optical and the cooling frequency, $\nu_c$, above
the optical 0.87 days after the burst.

     This conclusion is supported by the detailed modelling of
\citet{KP2003}.  They found that the early- and late-time optical data
is consistent with a low-density, homogeneous external medium and
inconsistent with a wind-stratified external medium.  Our broad band
spectral modelling supports this conclusion.  However, we are unable
to rule out the possibility that the burst expanded into a shocked
wind, or into an environment where the equatorial wind is
significantly stronger than the polar wind.

     If the burst did expand into a homogeneous medium then the
observed rate of decay of the OA at $t = 0.87$ days ($\alpha = 1.01
\pm 0.02$) and case 3 of Table~\ref{TABLE:predict} implies that the
electron index is $2.3 \pm 0.1$ and the intrinsic spectral slope at
this time is $0.67 \pm 0.01$.  This is consistent with the dereddened
spectral slope at this time ($\beta = 0.69 \pm 0.14$) that we
determined in \S~\ref{SECTION:sed}.  Eq.~11 of \citet{SPN1998} implies
that the cooling frequency at $t = 0.87$ days in a homogeneous medium
is $\nu_c = 2.9 \times 10^{12} \epsilon_B^{-3/2} n^{-1}$ Hz.  For the
case of a homogeneous ISM \citet{KP2003} find $\epsilon_{B_f} \approx
0.001$ in the forward shock at 11 minutes and an environmental density
of $10^{-3} \la n \la 10^{-2}$ cm$^{-3}$.  Using these value we
predict that the cooling frequency at $t = 0.87$ days should be
approximately $9 \times 10^{18} \la \nu_c \la 9 \times 10^{19}$ Hz,
which is in the high-energy regime (approximately 37--370 keV).
However, the modeling of \citet{KP2003} suggests that $\epsilon_{B_f}$
decreases by a factor of two between eleven minutes and ten days after
the burst, so locating the location of the cooling frequency is
somewhat uncertain.  Regardless, the \citet{KP2003} model is
inconsistent with $\nu_c$ being near the optical at this time.
Therefore, we do not believe that the cooling break moved through
optical frequencies between 0.1 and 1.0 days after the burst.

     The emerging picture of the evolution of this burst is a
transition from the emission being dominated by the reverse shock to
emission being dominated by the forward shock at approximately five
minutes after the burst.  The reverse shock has an optical decay of
$\alpha_1 = 1.97 \pm 0.05$, the forward shock has an optical decay of
$\alpha_2 = 1.01 \pm 0.02$ and the spectral index of the forward shock
photons is $\beta = 0.69 \pm 0.14$.  The spectral and decay slopes
imply that the electron index is $p = 2.3 \pm 0.1$ and are consistent
with the local environment dominated by a homogeneous ISM\@.


\section{Discussion\label{SECTION:discussion}}

\subsection{A Comparison with GRB~990123\label{SECTION:grb990123}}

     The overall shape of the optical decay of \objectname{GRB~021211}
appears to be very similar to that of \objectname{GRB~990123} (\eg,
\citealt{HBH2000}).  To quantify this we fit
Eq.~\ref{EQUATION:double_power_law} to the early-time $R_C$-band data
from \citet{FTM1999} and the data used by \citet{HBH2000}.  The best
fit occurs with power law slopes of $\alpha_1 = 2.25 \pm 0.48$ and
$\alpha_2 = 1.15 \pm 0.05$, a transition time of $t_t = 1.3 \pm 1.0$
minutes, and an $R_C$-band magnitude at transition of $R_C(t_t) =
10.91 \pm 1.33$ mag.  The goodness of fit was $\chi^2_{22} = 0.4335$.
The magnitude of the host was fixed at $R_C = 24.07$ mag
\citep{HBH2000} since we are interested in determining the slopes and
transition time under the assumption that there are two mechanisms
(the forward and reverse shocks) contributing to the flux, not the
magnitude of the host.

     The OA of \objectname{GRB~990123} made the transition from being
dominated by the reverse shock to being dominated by the forward shock
faster than the OA of \objectname{GRB 021211} did.  The rest-frame
transition times are 2.7 minutes for \objectname{GRB~021211} and 0.5
minutes for \objectname{GRB~990123}.  However, this difference may be
nothing more than an artifact of the data.  The
\objectname{GRB~990123} data consists of 26 data points from several
sources, so there may be a systematic offsets in the photometric
calibrations.  If we force the \objectname{GRB~990123} fit to have a
rest-frame transition time of 2.7 minutes (the same as
\objectname{GRB~021211} does) then the best fit double power law has
$\alpha_1 = 1.79 \pm 0.10$ and $\alpha_2 = 1.06 \pm 0.02$ with a
goodness of fit of $\chi^2_{22} = 0.5143$.  This is only marginally
worse than the best fit with the transition time is a free parameter
and the decay slopes are only $1.7 \sigma$ different from the fit to
the \objectname{GRB~021211} data.  This suggests that the overall
temporal evolution of the optical decays of both bursts was the same
to within the limitations of the data.

     These two OAs differ in that \objectname{GRB~021211} shows strong
evidence for rapid variations about a power law decay during the first
day while \objectname{GRB~990123} exhibits no variations
\citep{HBH2000}.  This may be an artifact of the sampling rate of the
\objectname{GRB~990123} data, or cross-calibration issues.  However, f
the smooth optical decay of \objectname{GRB~990123} is real then it
may indicate that the circumburst medium around
\objectname{GRB~990123} is smoother than that around
\objectname{GRB~021211} \citep{WL2000}.  Alternately,
\objectname{GRB~021211} may have had additional energy added via
refreshed shocks \citep{PMR1998} while \objectname{GRB~990123} did
not.

\begin{figure}
\plotone{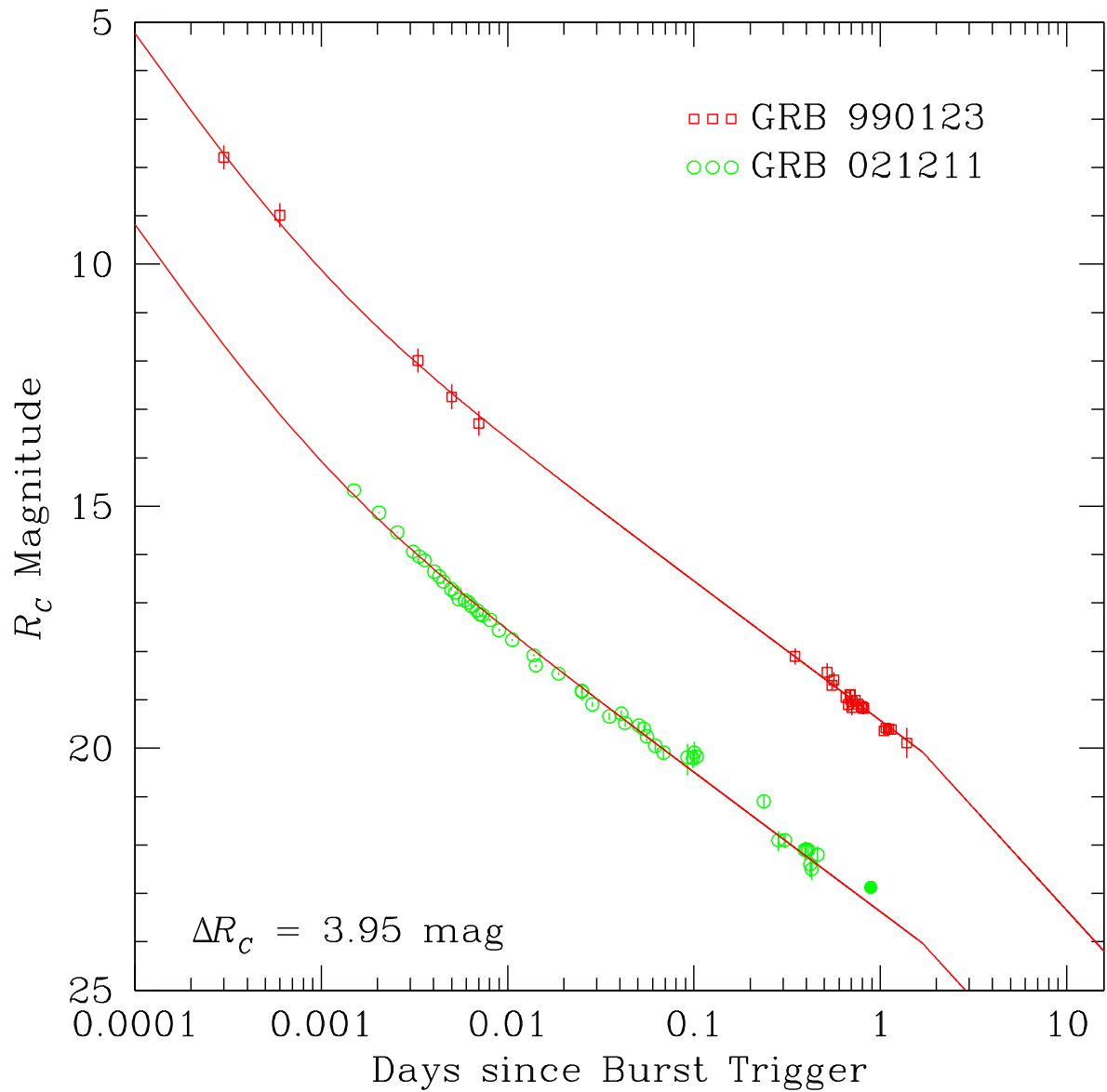} \figcaption[./compare.eps]{This Figure shows
the \protect\objectname{GRB~990123} $R_C$-band data (open squares)
with the best-fitting double power law as described in
\S~\ref{SECTION:grb990123}. The published
\protect\objectname{GRB~021211} $R_C$-band data (open circles), and
our $R_C$-band data (closed circles), is also shown.  The best-fitting
double power law to the \protect\objectname{GRB~990123} $R_C$-band
data superimposed has been shifted 3.95 mag fainter so that it
overlies the \protect\objectname{GRB~021211} data.
\label{FIGURE:compare}}
\end{figure}

     Figure~\ref{FIGURE:compare} shows the $R_C$-band data for
\objectname{GRB~990123} with the best-fitting double power law
superimposed.  The poor temporal coverage of the data makes it
difficult to determine the transition time.  This Figure also shows
the $R_C$-band data for \objectname{GRB~021211} with the fit for
\objectname{GRB~990123} superimposed.  With the exception of the
short-timescale fluctuations the shape of the \objectname{GRB~990123}
fit is in good agreement the optical decay of \objectname{GRB~021211}.
The data for \objectname{GRB~990123} has been adjusted to correct for
the difference in redshift between the two GRBs.  The fluxes with
shifted to match the observed $R_C$ band of \objectname{GRB~021211} by
assuming a power law spectrum with $\beta = 0.75$ \citep{HBH2000} for
\objectname{GRB~990123}.  Next the difference in distance modulus
($\Delta \mu = 1.25$ mag) was applied.  Finally, a $K$ correction was
applied to compensate for the difference in redshift between the two
bursts.  After these effects had been taken into account the OA of
\objectname{GRB~021211} was 3.95 mag fainter than the OA of
\objectname{GRB~990123}, which corresponds to a difference in
intrinsic luminosity of 38 times.  Except for the small-timescale
fluctuations this difference is approximately constant between
approximately 90 seconds and one day after the burst.

     The spectra of these two OAs are similar.  \citet{HBH2000} used
broadband photometry to find $\beta = 0.75 \pm 0.07$ for
\objectname{GRB~990123} between zero and three days after the burst.
This is almost identical to the value for \objectname{GRB~021211}.
\citet{ACH1999} found $\beta = 0.69 \pm 0.10$ from a spectrum taken 48
minutes after the burst.  The similarity in the spectra of the two
bursts implies that they had similar electron indices.
 

\subsection{The Host Galaxy\label{SECTION:host}}

     The location of the OA in our 2003 Jan.\ 29 UT WIYN image was
determined by a comparison with {\sl HST\/}/ACS/WFC images taken on
2003 Dec.\ 24/25 UT \citep{FLV2002}.  Our late-time image has a point
source with $R_C = 25.07 \pm 0.12$ mag at the location of the OA\@.
\citet{FLV2002} find that this source is resolved with an intrinsic
full width at half maximum of $\approx0\farcs13$ ($\approx 1$ proper
kpc) once the point-spread function of the ACS is removed.  This is
beyond the ability of WIYN to resolve so it appears as a point source
in our data.  The {\sl HST\/}/ACS/WFC data give an AB magnitude in the
F606W filter of $25.3 \pm 0.2$ mag, which corresponds to $R_C \approx
25.1$ mag.  We find $R_C = 24.93 \pm 0.17$ mag at $t = 21.6$ days
after the burst and $R_C = 25.07 \pm 0.12$ mag at 48.8 days.

     The $V\!-\!R$ colour of the afterglow at $t = 21.6$ days is $0.47
\pm 0.23$ mag, after correcting for Galactic reddening, which
corresponds to a spectral slope of $\beta = -1.4 \pm 1.2$.  The data
in \citet{RC3} suggest that the spectral slope of the Small Magellanic
Cloud (SMC) at near-ultraviolet wavelengths is $\beta_\mathrm{SMC}
\approx -1.6$.  This is consistent with the observed colour of the OA
at similar rest-frame wavelengths 21.6 days after the burst.

     However, \citet{DBM2003} showed that the OA has a supernova
component that reached its maximum light at $\approx 25$ days after
the burst.  They find that, at this time, the host and the supernova
contribute approximately equally to the light while the OA contributes
less than approximately 5\% of the light.  Their best estimate of the
host magnitude is $R_C = 25.22 \pm 0.10$ mag.  The corresponding
absolute rest-frame magnitude of the host is $M_U \approx -19$ mag.

     \citet{LTH1995} find a typical magnitude of
${(M^*_B)}_\mathrm{AB} = -21.4$ mag for blue galaxies at $0.75 \le z
\le 1.00$ if $(H_0,\Omega_m,\Omega_\Lambda) = (50,1,0)$.  For our
adopted cosmology this corresponds to ${(M^*_B)}_\mathrm{AB} = -22.2$
mag.  If we assume that the host galaxy has a power-law spectrum with
$\beta_\mathrm{gal} = -1.4$ (as implied by the $V\!-\!R$ colour of the
afterglow at $t = 21.6$ days) then the rest-frame $B$-band luminosity
is $L_B \approx 0.1 L^*_B$ where $L^*_B$ is the rest-frame $B$-band
luminosity of a typical blue galaxy at $z \approx 1$.  Therefore the
host of \objectname{GRB~021211} appears to be subluminous.  This is
consistent with \citet{HF1999} who found that GRB host galaxies tend
to be subluminous.  Our comparison of the total $B$-band luminosity of
the host to $L^*_B$ is somewhat uncertain since $M^*$ is highly
correlated with the slope of the faint end of the galaxy luminosity
function as well as with its normalization \citep{LTH1995}.

     At a redshift of $z = 1.004$ a rest-frame wavelength of 2800
{\AA} corresponds to the observer's $V$ band.  Therefore we are able
to use Eq.~2 of \citep{MPD1998} to estimate the integrated star
formation rate in host galaxy from the rest-frame continuum flux at
2800 {\AA}.  We find a star-formation rate of $\approx 1$
$\mathcal{M}_\sun \mathrm{yr}^{-1}$.  This calculation depends on the
details of the initial mass function, and it assumes that there is no
extinction in the host galaxy.  We know from the observed SED of the
OA (see Sect.~\ref{SECTION:sed}) that there is negligible extinction
in the host along the line of sight to the burst.  However we do not
know what level of extinction is present in the rest of the host.
Therefore we estimate that the integrated star-formation rate in the
host galaxy of \objectname{GRB~021211} is $\ga 1$ $\mathcal{M}_\sun
\mathrm{yr}^{-1}$.

     The specific star-formation rate per unit blue luminosity of the
host galaxy of \objectname{GRB~021211}, if we ignore extinction within
the host, is $\approx 10$ $\mathcal{M}_\sun \mathrm{yr}^{-1}
{L^*_B}^{-1}$, which is similar to that of other GRB host galaxies.
Our imaging data do not have high enough resolution to determine
morphological properties of the host.  That will require {\sl HST\/}
imaging of the host after the OA has faded.  However the early {\sl
HST\/}/ACS imaging of \citet{FLV2002} suggests that the host has a
diameter of $\approx 1$ proper kpc.  This implies a star-formation
rate density of $\ga 1.3 \mathcal{M}_\sun \mathrm{yr}^{-1}
\mathrm{kpc}^{-2}$ in the host.  This is at the low end of the range
of values (1--1000 $\mathcal{M}_\sun \mathrm{yr}^{-1}
\mathrm{kpc}^{-1}$) that characterize a starburst galaxy
\citep{K1998}.


\section{Conclusions\label{SECTION:conc}}

     We present $BV\!{R_C\!}J\!H{K_s}$ photometry of the OA of
\objectname{GRB021211} taken at the Magellan, MMT, and WIYN
observatories.  These data were taken between 0.7 and 50 days after
the burst.  The broad-band optical/infrared SED yields an intrinsic
spectral slope of $0.69 \pm 0.14$ at 0.87 days after the burst and we
find no evidence for colour evolution in the optical between 0.1 and
1.0 days.  There is weak evidence for brightness fluctuations in the
$J$ and $K_s$ bands approximately 0.7 days after the burst.  We find
no evidence for extragalactic extinction along the line of sight to
the burst.  The optical light is dominated by power law decay with an
index of 1.97 until 5.46 minutes after the burst.  After this time the
decay is dominated by a power law with an index of 1.01.  This is
consistent with the early time flux being dominated by a reverse shock
and a transition to domination by the forward shock 5.46 minutes after
the burst.

     There is evidence for rapid fluctuations in the flux about the
smooth power-law decay similar to what has been seen in
\objectname{GRB~011211} \citep{HSG2002}, \objectname{GRB~021004}
\citep{BSW2003,MGF2003}, and \objectname{GRB~030329} \citep{MGS2003}.
These fluctuations may also include colour variations, but there is
insufficient data to determine this is a convincing way.

     The spectral slope has been combined with the observed $R_C$-band
optical decay to determine that the shocked electrons are probably in
the slow cooling regime with an electron index of $2.3 \pm 0.1$, and
that the burst probably occurred in a homogeneous ISM\@.  There is
weak evidence for a jet break between 0.89 and 1.87 days after the
burst, and \citep{KP2003} suggest that the jet break may have occurred
before ten days based on the radio flux at 8.5 GHz.  If the jet break
did occur before ten days then the half-opening angle of the jet is
$1\fdg4 < \theta_j < 4\fdg4$, which is one of the smallest opening
angles of any GRB jet \citep{BFK2003}.  These angles imply that the
total gamma-ray energy in the burst was $3 \times 10^{48}$ erg $ <
E_\gamma < 3 \times 10^{49}$ erg, which is much less than the
canonical value of \citet{BFK2003} yet similar to that inferred from
the first jet break in \objectname{GRB~030329}.  This suggests that
most of the energy in \objectname{GRB~021211} may be not be in the OA,
but in a different component such as a frozen in magnetic field (cf
\citealt{KP2003}), kinetic energy in the supernova ejecta, or a second
jet component.

     The optical decay of \objectname{GRB~021211} during the first day
after the burst was very similar to that of \objectname{GRB~90123},
only the OA of \objectname{GRB~021211} was instrinsically $\approx 38$
times fainter.  Both OAs had an initial steep decline with a power law
index of $\approx 2$.  This is consistent with the OA being dominated
by emission from a reverse during the first few minutes after the
burst.  The OA of \objectname{GRB~990123} made the transition from
being dominated by the light from the reverse shock to being dominated
by the light from the forward shock $1.3 \pm 1.0$ minutes after the
burst (0.5 minutes in the rest frame of \objectname{GRB~990123}).  The
transition occurred at $5.46 \pm 0.80$ minutes for
\objectname{GRB~021211} (2.7 minutes in the rest frame).  After the
transition each OA is dominated by forward shock emission with decay
indices of $\approx 1$.  The similarity in the decays of these two OAs
suggests that the environments and physics of \objectname{GRB~990123}
and \objectname{GRB~021211} were similar.

     The rapid localization of \objectname{GRB~021211} and the
near-continuous monitoring of its OA during the first day after the
burst occurred has allowed the early time evolution of this burst to
be studied in detail.  This detailed coverage of the optical decay has
made it possible to recognize the similarity between
\objectname{GRB~021211} and \objectname{GRB~990123}.  The OA of this
burst faded very rapidly for the first several minutes, and had the
initial afterglow searches been delayed by even one hour the OA would
have been $\approx 5$ mag fainter.  It is questionable whether or not
the OA would have been discovered in that case.  If the OA had not
been detected rapidly this would have been classified as a dark burst.
Further, without the prompt discovery of the OA the rapid early time
decay would not have been identified.  Unfortunately there is very
little data for the period between one and 20 days after the burst, so
it is not possible to determine when the jet break occurred.  Daily
monitoring of OAs during the first few weeks will be possible using
the the Ultraviolet/Optical Telescope on the upcoming {\sl Swift\/}
mission.  This will monitoring will make it much less likely that
critical periods of the decay, such as the jet break, will be missed
for bursts that occur during the {\sl Swift\/} era.


\acknowledgements

     We wish to thank the {\sl HETE-II\/} team, Scott Barthelmy, and
the GRB Coordinates Network (GCN) for rapidly providing precise GRB
positions to the astronomical community.  We also wish to thank Arne
Henden for providing precision photometry of stars in the field of
\objectname{GRB~021211}.  STH and PMG acknowledge support from the NASA LTSA
grant NAG5--9364.  This research has made use of the NASA/IPAC
Extragalactic Database (NED), which is operated by the Jet Propulsion
Laboratory, California Institute of Technology, under contract with
NASA\@.


  
\end{document}